%% file: main.tex
\documentclass[iop, revtex4]{emulateapj}
\usepackage{natbib}
\usepackage{lscape}
\usepackage{wasysym}
\usepackage{hyperref}
\hypersetup{
  colorlinks   = true,    
  urlcolor     = blue,    
  linkcolor    = blue,    
  citecolor    = blue      
}

\usepackage{graphicx}
\usepackage{float}

\shorttitle{Occurrence of giant planets in dusty systems}
\shortauthors{Meshkat et al.}

\begin{document}

\title{A Direct Imaging Survey of Spitzer detected debris disks:\\ Occurrence of giant planets in dusty systems$^{*}$}
\author{Tiffany Meshkat\altaffilmark{1,2}, 
Dimitri Mawet\altaffilmark{2,3},
Marta Bryan\altaffilmark{3},
Sasha Hinkley\altaffilmark{4},
Brendan P. Bowler\altaffilmark{5,$\dagger$},
Karl R. Stapelfeldt\altaffilmark{2,6},
Konstantin Batygin\altaffilmark{3},
Deborah Padgett\altaffilmark{2,6},
Farisa Y. Morales\altaffilmark{2},
Eugene Serabyn\altaffilmark{2},
Valentin Christiaens\altaffilmark{7,8,9},
Timothy D. Brandt\altaffilmark{10,$\ddagger$},
Zahed Wahhaj\altaffilmark{11}
}

\altaffiltext{1}{IPAC, Caltech, M/C 100-22, 1200 E. California Blvd, Pasadena, CA 91125, USA}
\altaffiltext{2}{Jet Propulsion Laboratory, California Institute of Technology, 4800 Oak Grove Drive, Pasadena, CA 91109, USA}
\altaffiltext{3}{California Institute of Technology, 770 S Wilson Ave, Pasadena CA 91125, USA}
\altaffiltext{4}{University of Exeter, Physics Department, Stocker Road, Exeter, EX4 4QL, UK}
\altaffiltext{5}{McDonald Observatory and the University of Texas at Austin, Department of Astronomy, 2515 Speedway, Stop C1400, Austin, TX 78712, USA}
\altaffiltext{6}{Laboratory for Exoplanets and Stellar Astrophysics, Code 667, NASA Goddard Space Flight Center, Greenbelt, MD 20771, USA}
\altaffiltext{7}{Departamento de Astronom\'ia, Universidad de Chile, Casilla 36-D, Santiago, Chile}
\altaffiltext{8}{Space sciences, Technologies \& Astrophysics Research (STAR) Institute, Universit\'e de Li\`ege, All\'ee du Six Ao\^ut 19c, B-4000 Sart Tilman, Belgium}
\altaffiltext{9}{Millenium Nucleus ``Protoplanetary Disks in ALMA Early Science'', Chile}
\altaffiltext{10}{Astrophysics Department, Institute for Advanced Study, Princeton, NJ, USA}
\altaffiltext{11}{European Southern Observatory, Alonso de C\`ordova 3107, Vitacura, Casilla 19001, Santiago, Chile}
\altaffiltext{$\dagger$}{Hubble Fellow}
\altaffiltext{$\ddagger$}{Sagan Fellow}
\altaffiltext{$^{*}$}{Some of the data presented herein were obtained at the W.M. Keck Observatory, which is operated as a scientific partnership among the California Institute of Technology, the University of California and the National Aeronautics and Space Administration. The Observatory was made possible by the generous financial support of the W.M. Keck Foundation.}

\begin{abstract}
We describe a joint high contrast imaging survey for planets at Keck and VLT of the last large sample of debris disks identified by the Spitzer Space Telescope. No new substellar companions were discovered in our survey of 30 Spitzer-selected targets. We combine our observations with data from four published surveys to place constraints on the frequency of planets around 130 debris disk single stars, the largest sample to date. For a control sample, we assembled contrast curves from several published surveys targeting 277 stars which do not show infrared excesses. We assumed a double power law distribution in mass and semi-major axis of the form f(m,a) =  $Cm^{\alpha}a^{\beta}$, where we adopted power law values and logarithmically flat values for the mass and semi-major axis of planets. We find that the frequency of giant planets with masses 5-20 $M_{\rm Jup}$ and separations 10-1000 AU around stars with debris disks is 6.27\% (68\% confidence interval 3.68 - 9.76\%), compared to 0.73\% (68\% confidence interval 0.20 - 1.80\%) for the control sample of stars without disks. These distributions differ at the 88\% confidence level, tentatively suggesting distinctness of these samples. 
\end{abstract}


\section{Introduction}

High angular resolution observations utilizing adaptive optics (hereafter ``AO'') and coronagraphy allow the study of exoplanets at separations of tens to hundreds of AU, outside of the reach of the transit and radial velocity detection methods.  Just as the radial velocity technique revealed an unexpected reservoir of planets in extremely close orbits around their stars, the high-contrast images of HR 8799bcde, $\beta$ Pic b, HD 95086 b, HD 106906 b, 51 Eri b, and HIP 65426 b \citep{Marois10b, Lagrange10, Rameau13b, Bailey14, Macintosh15, Chauvin17}, have demonstrated that planets of several Jupiter masses can also exist at astonishingly large orbital distances: $\sim$650 AU in the case of HD 106906 b and up to $\sim2000$ AU for GU Psc b \citep{Naud14}. Understanding the true frequency of these objects at wider separations allows planet formation theorists and modelers to fill out the census of planetary mass companions and more fully characterize the orbital architecture of planetary systems. Directly imaging planets opens the door for subsequent spectroscopic study of the planets themselves  \citep{Bowler10b,Barman11,Konopacky13,Ingraham14,Macintosh15,Bonnefoy16}. 

Most of the aforementioned directly imaged planets orbit stars with bright debris disks, tenuous dust clouds formed from the ongoing collisions of circumstellar rocky or icy parent bodies. Indeed, based on the presence and structure of the $\beta$ Pic disk, the existence of the planet $\beta$ Pic b was predicted well before its discovery \citep{Smith84,Beust00,Lagrange09}.  Moreover, gravitational stirring by planetary mass companions serves as a driver for the collisional processes which lead to dust production. The presence of a bright debris disk is thus a likely indicator that one or more planetary mass companions are present. Studying the dynamical interactions between planetary mass companions and their debris disks, is an opportunity to better understand the exoplanetary system as whole \citep{Chiang09, Boley12}. These systems are dynamical laboratories where we can study the induced morphology of the disks based on the secular and resonant interactions with perturbing planets \citep{Kennedy14,Lee16, Nesvold16,Nesvold17}. The disk shapes and morphologies can help to constrain the masses of any perturbing companions \citep{Morrison16}.

\citet{Bowler16} performed a meta-analysis on 384 stars with published direct imaging data and found the occurrence rate of giant planets to be $0.6^{+0.7}_{-0.5}\%$, using hot-start planet cooling models. This analysis included all published single stars across the full spectral range, with no additional selection for the presence of a debris disk. 
In this work, we compile the largest survey thus far of debris disk-selected targets in order to set strict limits on the occurrence of giant planets around stars with debris disks, as compared with a large control sample of diskless stars. We describe our observing campaign targeting the last significant sample of stars with debris disks from the Spitzer Space Telescope. We combine these data with four published deep high contrast imaging surveys of stars with debris disks.

In Section \ref{sec:target_sample}, we discuss our target sample selection,  observations, data reduction, and present the contrast limits achieved in our Keck and VLT data.
In Section \ref{sec:completed}, we combine our sample with contrast limits from three published surveys \citep{Wahhaj13,Rameau13,Janson13,Meshkat15} and describe the observing strategies from those surveys. We also discuss our selection criteria for the control sample of stars without debris disks.
In Section \ref{sec:results}, we explain our strategy for single epoch detections, we derive disk properties for our targets based on SED modeling, and we describe how we measure the companion occurrence rate. In Section \ref{sec:discussion}, we discuss how the companion occurrence rates differ between the debris disk and control samples and analyze the planet frequency in the context of our derived disk properties.

\section{Spitzer selected targets}
\label{sec:target_sample}

The Spitzer Space Telescope has conducted several surveys for debris disks around nearby stars. Although debris disks are more common around younger A stars \citep{Rieke05, Su06}, about 10\% of F, G, and K stars show signs of debris disks \citep{Trilling08, Carpenter09}, and these disks are rarer around M stars \citep{Gautier07}. 
The presence of a 22 or 24 micron excess in FGK type stars can help to rule out an advanced age for the system (i.e. it is unlikely that an FGK star with a significant excess will be older than $\sim$1 Gyr). Thus, for A and FGK stars, warm dust may be an indication of youth.  

Results for $\sim$600 targets are reported in the above studies, and represent the output of the Spitzer Legacy and GTO programs that were defined early in the mission. The largest remaining Spitzer debris disk survey is a volume-limited study of 600 additional F,G, and K stars within 25 pc of the Sun by \citep{Koerner10}. \citet{Koerner10} identified an additional 49 nearby stars with debris disks detected as infrared excess at either 24 or 70 $\mu$m.  These disks have characteristic fractional infrared luminosities of 0.01 \%, and radii of ~25 AU that cannot be spatially resolved by Spitzer.

In this work, we observed the 23 stars with the brightest infrared excesses with Keck/NIRC2 and VLT/NACO (see Table \ref{table:data_our_sample}). We have also included 7 additional nearby stars with known, young ages found to have debris disks from \citet{Plavchan09}, bringing the total number of new targets we observed to 30.

Three more targets were observed at the VLT as part of this survey (HIP 58576, HIP 72848, HIP 74975), however the infrared excess observed in these targets with Spitzer was not confirmed with the WISE instrument \citep{Patel14}. Thus, we have removed these targets from the following analysis.

\citet{Koerner10} only provides ages for some of the stars in that sample. Where available, we used literature ages which used a variety of estimation methods including young moving group membership, isochrone fitting, Lithium abundances, and Ca H\&K line emission. For targets which had no literature age, we derived the \citet{Koerner10} ages from their Ca H\&K emission line strengths, using the formula from \citet{Mamajek08}. For the targets lacking Ca H\&K line strengths (HIP 25775, HIP 44295, HIP 77952), we calculated the ages using an empirical relation to convert the excess emission into an age \citep{Rieke05}.

\subsection{Observations}

Data were obtained for 17 targets in 2010 (Program ID: C256N2, PI: Sasha Hinkley) and one target in 2012 (Program ID: C248N2, PI: Heather Knutson) with the Keck/NIRC2 instrument \citep{Wizinowich13} 
and 16 targets in 2010 (Program ID: 085.C-0635(A), 086.C-0505(A), PI: Dimitri Mawet) at the Very Large Telescope (VLT)/UT4 with NACO \citep{Lenzen03,Rousset03}.

NACO data were obtained in $L'$-band with the classical Lyot $0\farcs7$ diameter coronagraph. NIRC2 data were obtained in $Kp$-band with the corona300 300 mas diameter) aperture coronagraph. All data were obtained in pupil tracking mode, in order to perform angular differential imaging \citep[ADI; ][]{Marois06} and PSF subtraction using principal component analysis \citep[PCA; ][]{Soummer12, Amara12} on the data. Table \ref{table:data_our_sample} contains details about the observations for each of our targets, including observation date, total integration time and on-sky rotation. The observing strategy varied from target to target depending on the conditions on the night. In general, we aimed to obtain more than 15$^{\circ}$ sky rotation in order to minimize self-subtraction of a potential companion in the post-processing. Some of the targets were observed both with NIRC2 and NACO, to follow-up potential companions. We also obtained short unsaturated off-axis images for photometric calibration for all targets.

\subsection{Data Reduction}

We subtracted the stellar PSF and speckles in our data by processing the centered data cubes with principal component analysis (PCA: \citealt{Soummer12, Amara12}). This algorithm utilizes the sky rotation around each target, due to observing in pupil tracking mode, in order to model and subtract the stellar PSF and residual speckles. We performed PCA subtraction over the full field of view of our targets, with the star masked out. We detected a few candidate companions, six of which were shown to be consistent with background objects (Figure \ref{fig:background}) and five were apparent binaries (Table \ref{table:binaries}). No new substellar companions were found in our data.

\begin{deluxetable}{l c c c c}
\tabletypesize{\tiny}
\tablecaption{Relative astrometry of detected binaries. All targets were \\observed with VLT/NACO, except for HIP 88745 with Keck/NIRC2.}
\tablewidth{0pt}
\tablehead{\colhead{Name} &  \colhead{Date} & \colhead{Sep ('')} & \colhead{P.A. ($^{\circ}$)} & \colhead{Delta mag}}
\startdata
HIP 44295B &  2010-04-19 & $5.13\pm0.03$ & $179.7\pm0.4$ & $3.66\pm0.5$ \\
HIP 58576B &  2010-04-22 & $1.13\pm0.02$ & $334.7\pm1.0$ & $6.12\pm0.5$\\
HIP 73633B &  2010-04-22 & $3.85\pm0.02$ & $5.6\pm0.4$ & $1.86\pm0.5$\\
HIP 73633C &  2010-04-22 & $3.96\pm0.02$ & $4.7\pm0.4$ & $2.29\pm0.5$\\
HIP 88745B &  2010-09-27 & $1.19\pm0.01$ & $314.1\pm0.5$ & $2.6\pm0.5$ 
\enddata
\label{table:binaries}
\end{deluxetable}

\input{targets_spitzer_sample.tex}

In order to determine the sensitivity of our data, we injected fake companions into the raw data before the image processing. Data from NACO/VLT and NIRC2/Keck were both obtained with coronagraphs, thus we could not use the star itself as a photometric reference PSF. We used an unsaturated, off-axis reference PSF as a photometric calibrator to create fake planets, generated from several reference PSFs for NACO/VLT and NIRC2/Keck individually. We scaled down the flux of the unsaturated PSF and injected fake companion point sources at the 20 $\sigma$ level at three different position angles before PCA processing. Depending on the instrument field-of-view, we ran PCA out to $3\farcs3$ for Keck/NIRC2 and $13''$ for VLT/NACO, measured the resulting signal-to-noise from the fake injected planet, and rescaled our results to 5$\sigma$. We took the average signal-to-noise of the fake planets for each of the three position angles in order to determine the 5$\sigma$ limit. We repeated this process at larger radii in order to map the whole field of view, increasing in steps of 3 resolution elements. We note that for the targets with very little sky rotation ($<10^{\circ}$), there is significant self-subtraction at small inner working angles with ADI analysis. We account for this self-subtraction by injecting fake companions into the raw data and measuring the resulting signal-to-noise of the point source after post-processing.
 
Table \ref{table:data_our_sample} lists our 5$\sigma$ contrast limits from $0\farcs25$ and $5\farcs0$ from the injected fake planet detection limits. The Keck datasets achieve, on average, better sensitivity due to their increased integration time, on-sky rotation, and the larger aperture size.

\section{Completed high contrast imaging surveys}
\label{sec:completed}
We aim to constrain the frequency of giant planets around stars with debris disks by combining our survey results with four samples of debris disk-selected targets, resulting in the largest sample of dusty debris disk stars in direct imaging thus far\footnote{Data from on-going large programs such as Morales et al. \textit{in prep}, Gemini Planet Imager Exoplanet Survey, and the SPHERE GTO are not included in this work as they are not yet published.}. We combine our Spitzer dusty debris disk sample (30 targets from Keck and VLT) with published surveys specifically targeting stars with debris disks: the Gemini NICI Planet-Finding Campaign \citep[57 targets, ][]{Wahhaj13}, the NACO Survey of Young Nearby Dusty Stars \citep[29 targets,][]{Rameau13}, the Strategic Exploration of Exoplanets and Disks with Subaru \citep[41 targets,][]{Janson13}, "Holey Debris Disks" survey (15 targets, \citealt{Meshkat15} and Bailey et al. \textit{in prep}). These targets increase the sample size and thus the statistical significance of our analysis. For the targets which have duplicate observations among the surveys, we used the more sensitive contrast curve at $1\arcsec$ in the subsequent analysis. Most contrast curves were presented as $5\sigma$ limits, except \citet{Wahhaj13} and \citet{Janson13} which we modified to be consistent with $5\sigma$ (detailed below). We note that, after subtracting off the stellar contribution using classical ADI and/or PCA, the distribution of noise in the image is approximately Gaussian \citep[see e.g.][]{Mawet14}. Thus, we have 130 individual stars in total (binaries have been removed). Table \ref{table:targets} shows the complete target list. The debris disk-selected surveys are discussed in detail below.

\subsection{The Gemini-NICI Planet-Finding Campaign}
\citet{Wahhaj13} used the Gemini/NICI instrument to search for exoplanets around young stars with debris disks. Targets were selected based on the presence of an infrared excess. Data were obtained in ADI mode with H and angular spectral differential imaging (ASDI) mode on and off the $CH_{4}$ narrow band. Several companion candidates were detected, most were shown to be consistent with background stars. Some of the companion candidates were not followed up and remain single epoch companion candidates. Using Bayesian analysis with flat priors, they find that less than 20\% of stars with debris disks have companions more massive than 3 $M_{\rm Jup}$ beyond 10 AU. They conclude that systems like $\beta$ Pic and HR 8799 are likely rare. Contrast curves are presented as a 95\% completeness threshold which corresponds to a source bright enough to be detected if it were located on 95\% of the background fluctuations, assuming a $3\sigma$ minimum for follow-up. Given that a Gaussian distribution with zero mean has 95\% of its probability above $1.64\sigma$, a $4.64\sigma$ source would be detected at $3\sigma$ or better, 95\% of the time. We therefore convert the 95\% limits to 5$\sigma$ by subtracting $2.5 \log (4.64/5) \approx 0.081$ mag, and proceed to include the modified sensitivity curves in the remainder of our analysis.

\input{debris_disk_sample_properties.tex}

\subsection{NaCo Survey of Young Nearby Dusty Stars}
The \citet{Rameau13} survey targeted young, nearby stars with dusty debris disks searching for giant planets. Data were obtained with the VLT/NACO instrument in $L'$-band. The presence of debris disks were inferred based on high infrared excesses in 24 and/or 70 $\mu m$. The HD 95086 b planet was discovered as part of this survey (reported in \citet{Rameau13b}). Following \citet{Bonavita12}, they find that the fraction of stars with giant planets (1-13 $M_{\rm Jup}$) at large separations (1-1000 AU) is 10.8\% to 24.8\%, at 68\% confidence level. This high fraction is likely due to the large bounds which include planet masses and separations to which these data are not sensitive (i.e. 1 $M_{\rm Jup}$ at 1 AU). Data are presented in 5 sigma contrast, and thus not rescaled for our meta-analysis.

\subsection{Strategic Exploration of Exoplanets and Disks with Subaru}
\citet{Janson13} use the Subaru/HiCIAO instrument to search for planets and detect scattered light from debris disk systems. Target selection was based on infrared excesses as well as inferred disk properties from SED modeling such as fractional luminosity and the approximate angular separation. Disks which show the SED signature of a spatially separated warm and cold disk were given priority in the sample. Several companion candidates were detected in this survey and shown to be consistent with background sources. Detection limits are presented at 5.5 sigma, which we modify to 5 sigma contrast by adding $2.5 \log (5.5/5) \approx 0.103$ mag.

\subsection{``Holey Debris Disks'' survey}
\citet{Meshkat15} and Bailey et al. (\textit{in prep}) obtained data on fifteen young stars which were selected based on membership in young moving groups, and bright debris disks for SED modeling. This project was dubbed the ``Holey Debris Disks'' survey, based on the holes or gaps in debris disks where the planets are expected to reside. HD 106906 b was discovered by this survey \citep{Bailey14} with Magellan AO + Clio2 system. The planetary mass companion ($11\pm2 M_{\rm Jup}$) was discovered at a projected separation of $7\farcs1$ (650 AU). Contrast limits are in 5 sigma, and thus not rescaled for meta-analysis.

\subsection{Control Sample}
\begin{figure}
\epsscale{1.0}
 \plotone{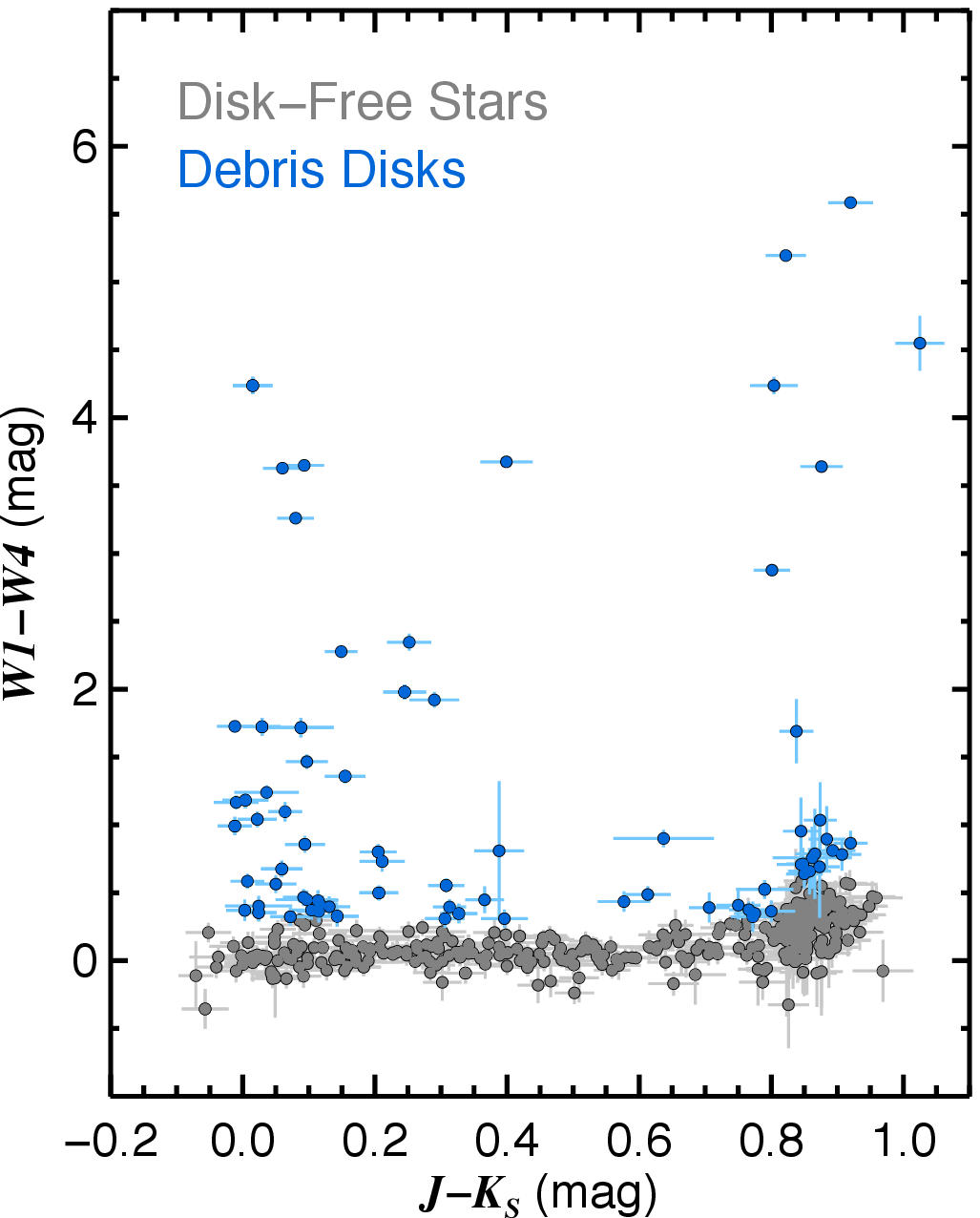}
 \caption{Vetting criteria for the targets included in our disk-free control sample. We included targets from several surveys \citep{Lafreniere07, Vigan12, Biller13, Nielsen13, Brandt14, Bowler15, Galicher16} and ruled out those with $W1-W4<0.3$ mag for $J-Ks <0.8$ mag, or $W1-W4<0.6$ mag for $J-Ks>0.8$ mag.}
 \label{fig:vetting}
\end{figure}

In order to assess the correlation of debris disks with giant, long period planets, we compiled a sample of stars without known debris disks to act as a control sample for our analysis. We included targets from several completed surveys searching for planets \citep{Lafreniere07, Vigan12, Biller13, Nielsen13, Brandt14, Bowler15, Galicher16}. In order to rule out targets with debris disks, we used several target selections. We removed all stars which overlapped with the targets in this survey, also removed targets from the \citet{Wahhaj13}, \citet{Rameau13}, \citet{Janson13}, or \citet{Meshkat15} surveys. We only included targets with $J$, $Ks$, $W1$, and $W4$ photometry. A target was considered ``diskless" if it had no W1-W4 excess. We used conservative cuts for excesses with a stepwise change at $\sim$K7: $W1-W4<0.3$ mag for $J-Ks <0.8$ mag, or $W1-W4<0.6$ mag for $J-Ks>0.8$ mag (see \autoref{fig:vetting}). Targets with binaries within 100 AU were also excluded from the target list. In total, we have 277 targets in our control sample. All contrast curves are scaled to 5 sigma contrast limit.

\begin{figure}
\epsscale{1.0}
 \plotone{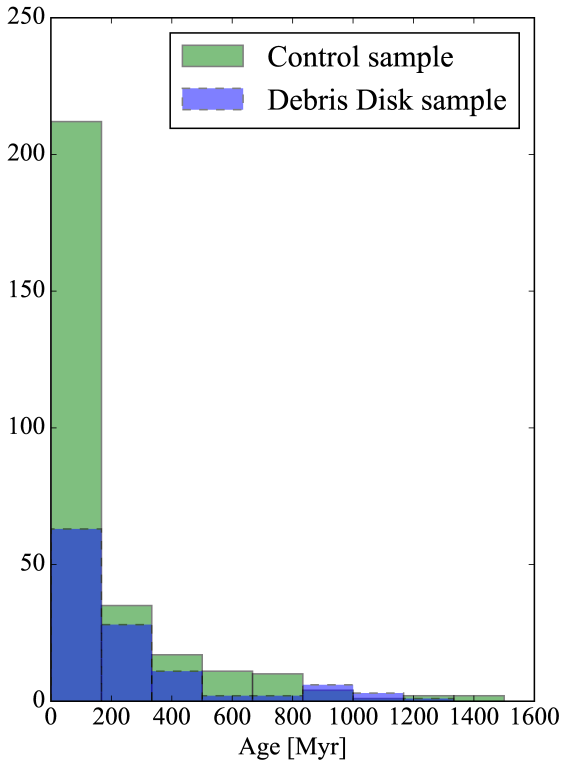}
 \caption{Age distribution for the targets in our debris disk sample (light blue) and the control sample (green). The overlapping regions between these targets is dark blue. }
 \label{fig:ages}
\end{figure}

\begin{figure}
\epsscale{1.1}
 \plotone{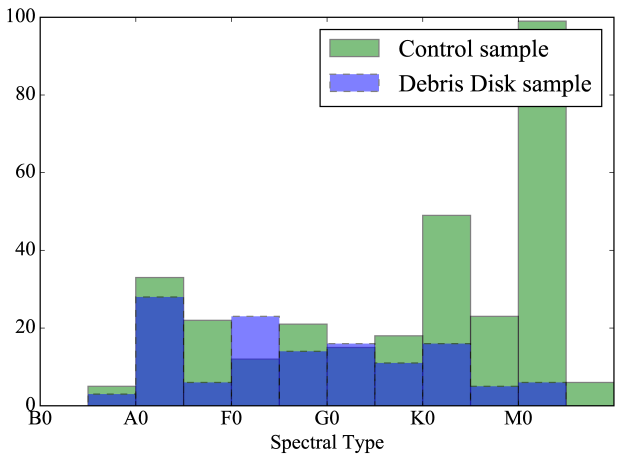}
 \caption{Spectral types for our debris disk sample (blue) and control sample (green).}
 \label{fig:sp_type}
\end{figure}

\autoref{fig:ages} and \autoref{fig:sp_type} compare the ages and spectral types of the debris disk sample with our control sample. We quantified the differences between these distributions by fitting the ages for each sample using a two Gaussian model.  We accounted for the uncertainties in the ages for each star by repeating this model fit 5,000 times, where each time we drew a random age from the Gaussian distribution for each age. This resulted in distributions of best fit heights, widths, and means for a two-Gaussian model of both the control and the debris disk populations. We then compared these distributions of six parameters using their Bayesian Information Criterion (BIC). BIC is defined as follows:  $BIC = -2 L + k \ln(n)$, where $L$ is the likelihood of the model, $k$ is the number of model free parameters, and $n$ is the number of data points.  The lower the BIC value, the better the model fit. Although the likelihood can be increased simply by fitting a more complicated model with more free parameters, BIC selects against these models with a penalty term. When comparing two models, if the delta BIC between them is $>$ 10, this is strong evidence that the model with the lower BIC value is a better fit \citep{Kass95}. For this model comparison, we fit the combined distributions between the debris disk and control sample populations for heights, widths, and means with a one Gaussian model and a two Gaussian model.  We found that for all six parameters the single Gaussian model was preferred (delta BIC $>$ 10), indicating that these age samples are drawn from the same underlying distribution. The increased number of M-stars in the control sample is discussed in Section \ref{sec:occurrence} below.

\section{Results}
\label{sec:results}

\subsection{Planetary mass companions}
\label{sec:results_pmcs}

In our debris disk sample, seven planetary mass companions ($<20\,M_{\rm Jup}$) were discovered or re-detected: HR8799 bcde \citep{Wahhaj13,Rameau13}, $\beta$ Pic b \citep{Wahhaj13,Rameau13}, HD 95086 b \citep{Rameau13b}, and HD 106906 b \citep{Bailey14}. The debris disk measurements for these targets are all well-studied and resolved (HR8799; \citealt{Matthews14, Booth16}, $\beta$ Pic; \citealt{Smith84}, HD 95086; \citealt{Su15,Moor13}, and HD 106906; \citealt{Chen05, Kalas15}).  For our statistical analysis, we consider these seven planetary mass companions to be four independent detections, as we treat the HR 8799 four planets as one planetary system detection.

In the diskless control sample, two planetary mass companion were detected: AB Pic B \citep{Chauvin05} and GJ 504 b \citep{Kuzuhara13}. AB Pic was included as a target in the \citep{Rameau13} dusty debris disk survey, based on the \citet{Zuckerman11} excess in the 12$\mu m$ IRAS and 24$\mu m$ MIPS/Spitzer channels, suggesting a warm belt. However, no Herschel cold excess was detected and this target has no IRS data. Based on our reanalysis of this target's SED we conclude that the excess from the warm belt is too tentative to be considered a robust detection of a debris disk. Given the non-detection with Herschel and uncertainty in the presence of a warm belt, we conservatively include AB Pic B in the diskless control sample. The low mass of the planet GJ 504 b was inferred based on the young age of the star ($\sim$160 Myr). However, recent reassessments of the age of GJ 504 suggests a much older age ($\sim$2.5 Gyr; \citealt{DOrazi16,Fuhrmann15}), and thus the companion is more likely to be a 30-40 Jupiter mass brown dwarf. Based on the age reassessment, we do not include this companion as a planetary mass detection in our subsequent statistical analysis. Thus, we have one planetary mass companion (5-20 $M_{\rm Jup}$) in the diskless control sample.

\subsection{Companion candidates and single epoch detections}

\begin{figure*}
 \epsscale{1.0}
 \plottwo{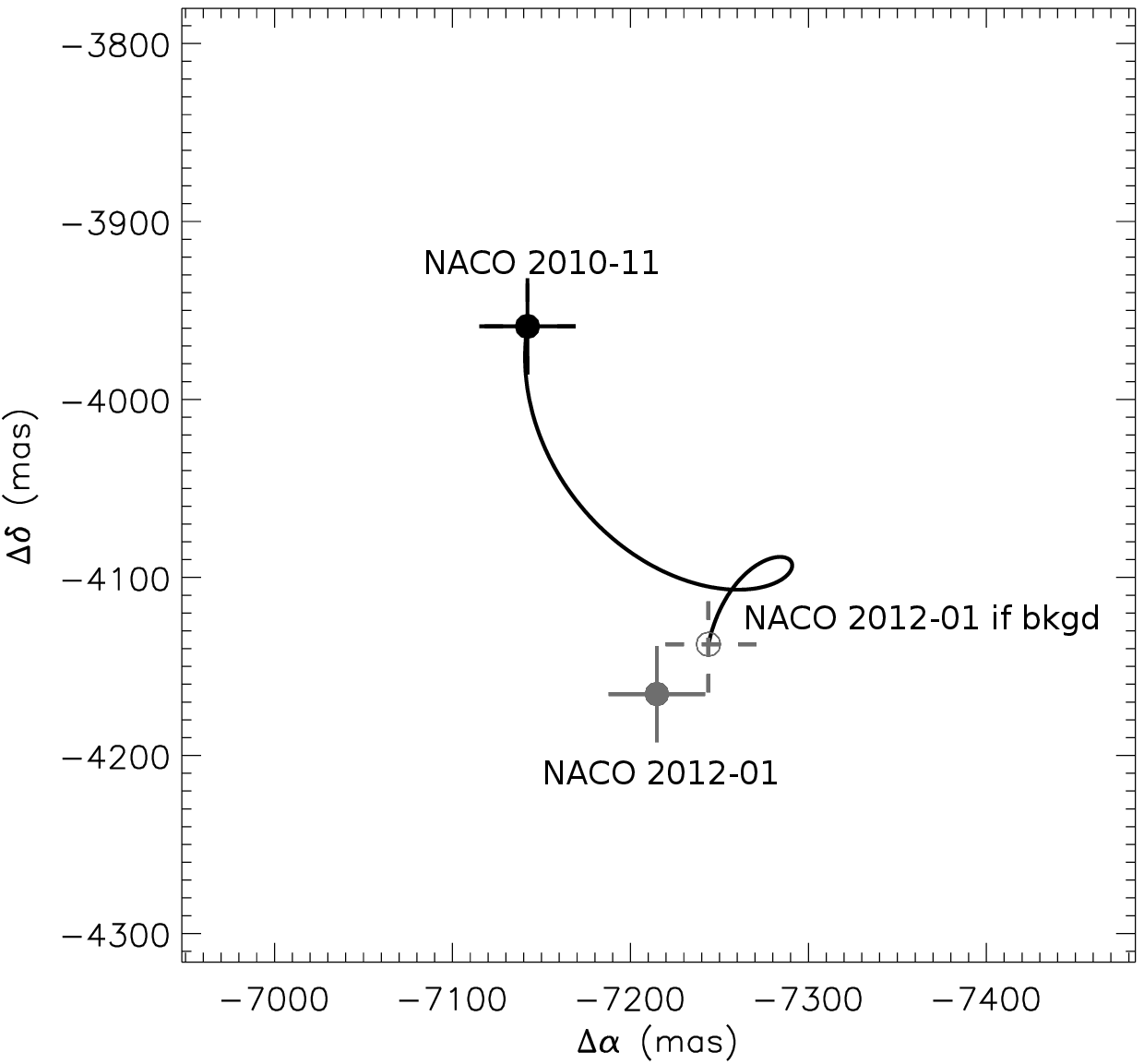}{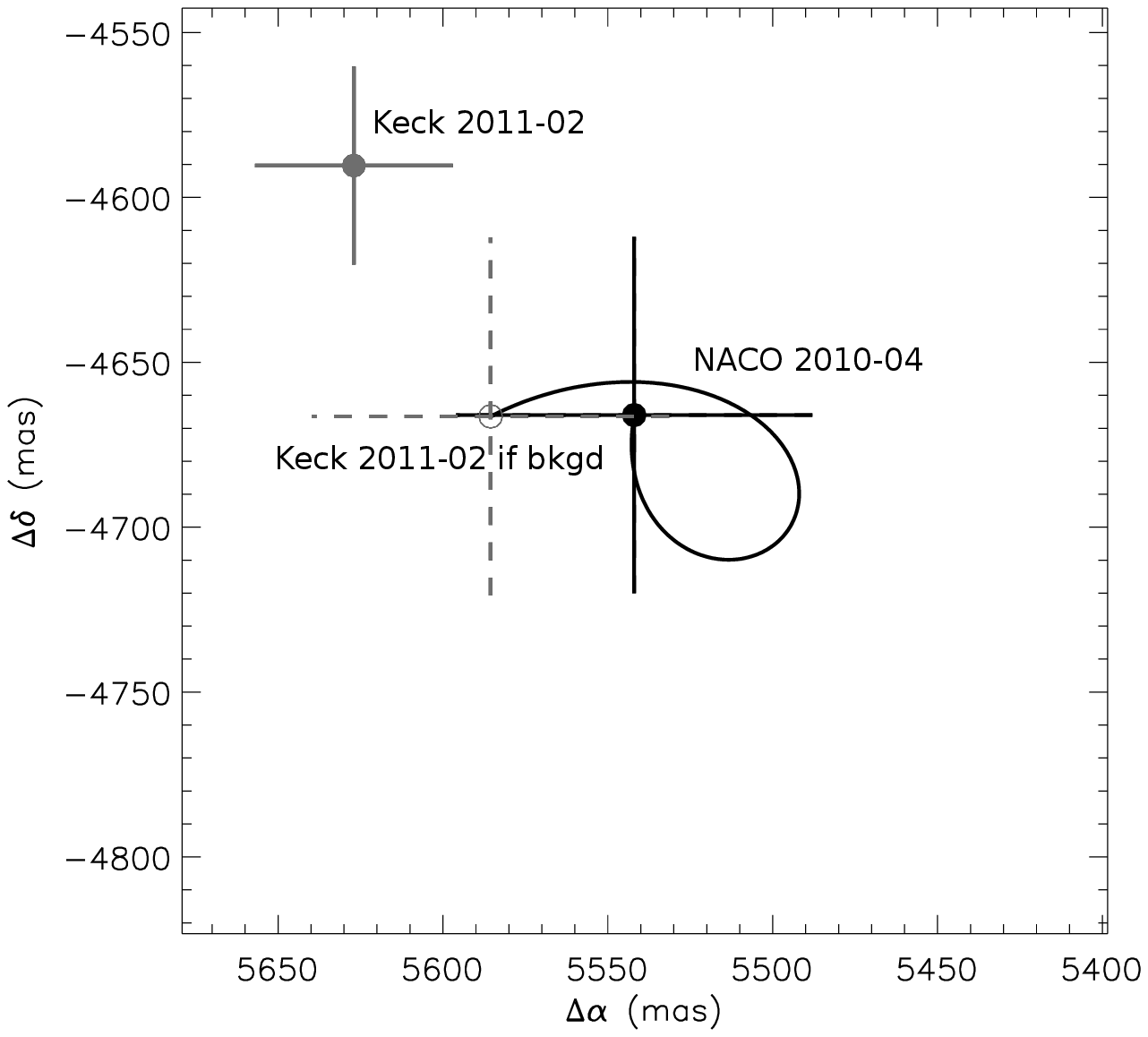}
 \plottwo{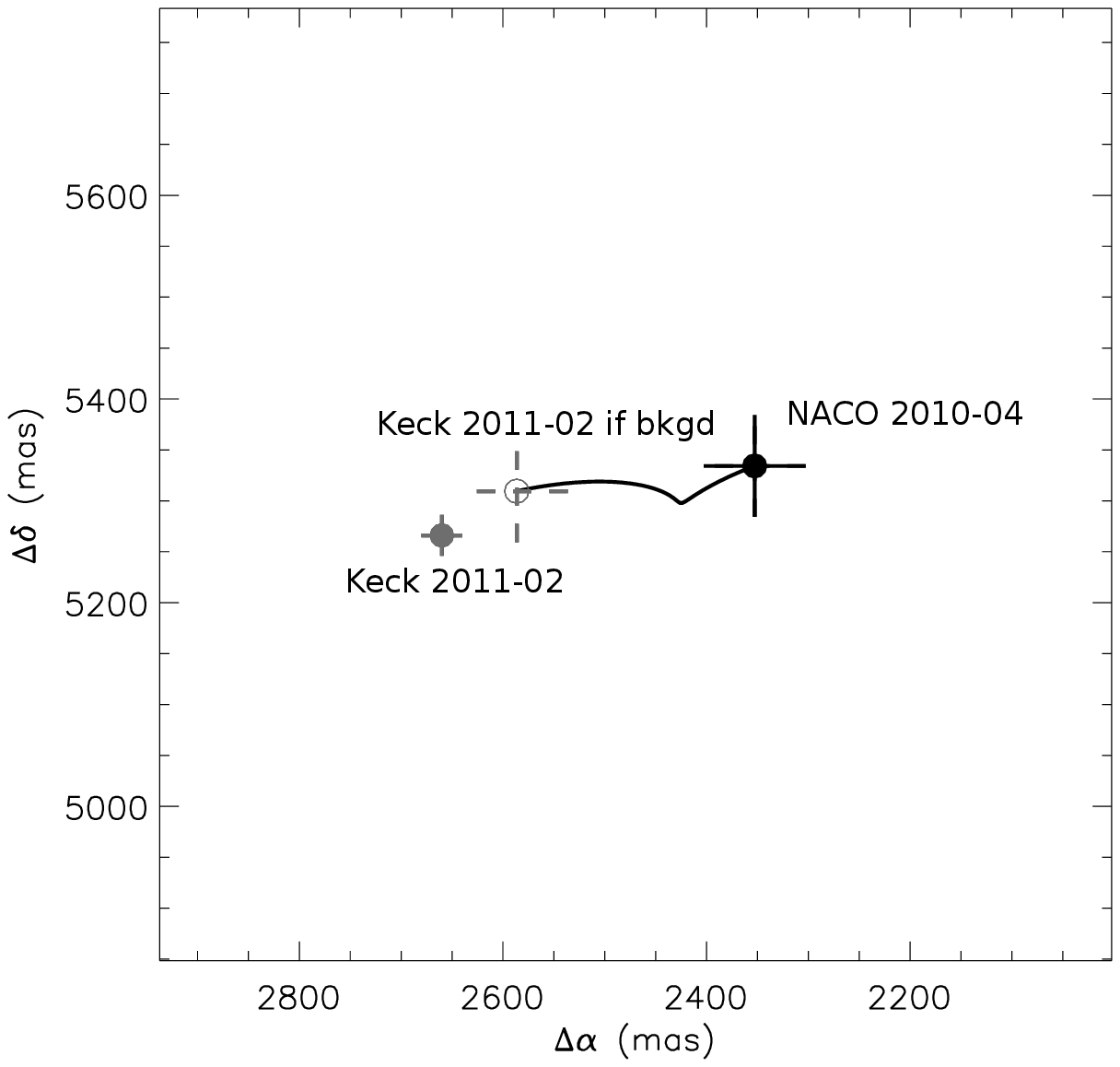}{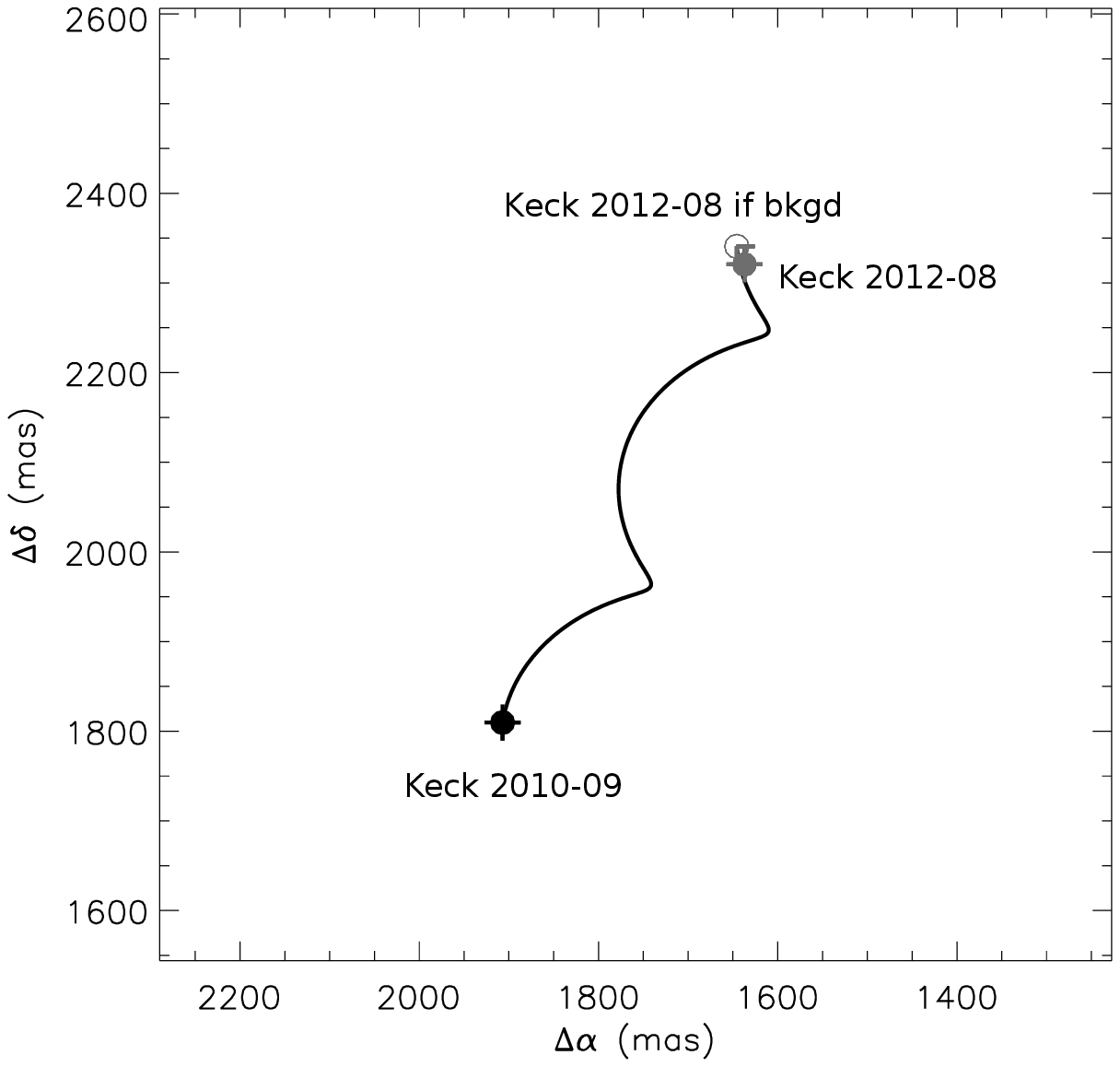}
 \plottwo{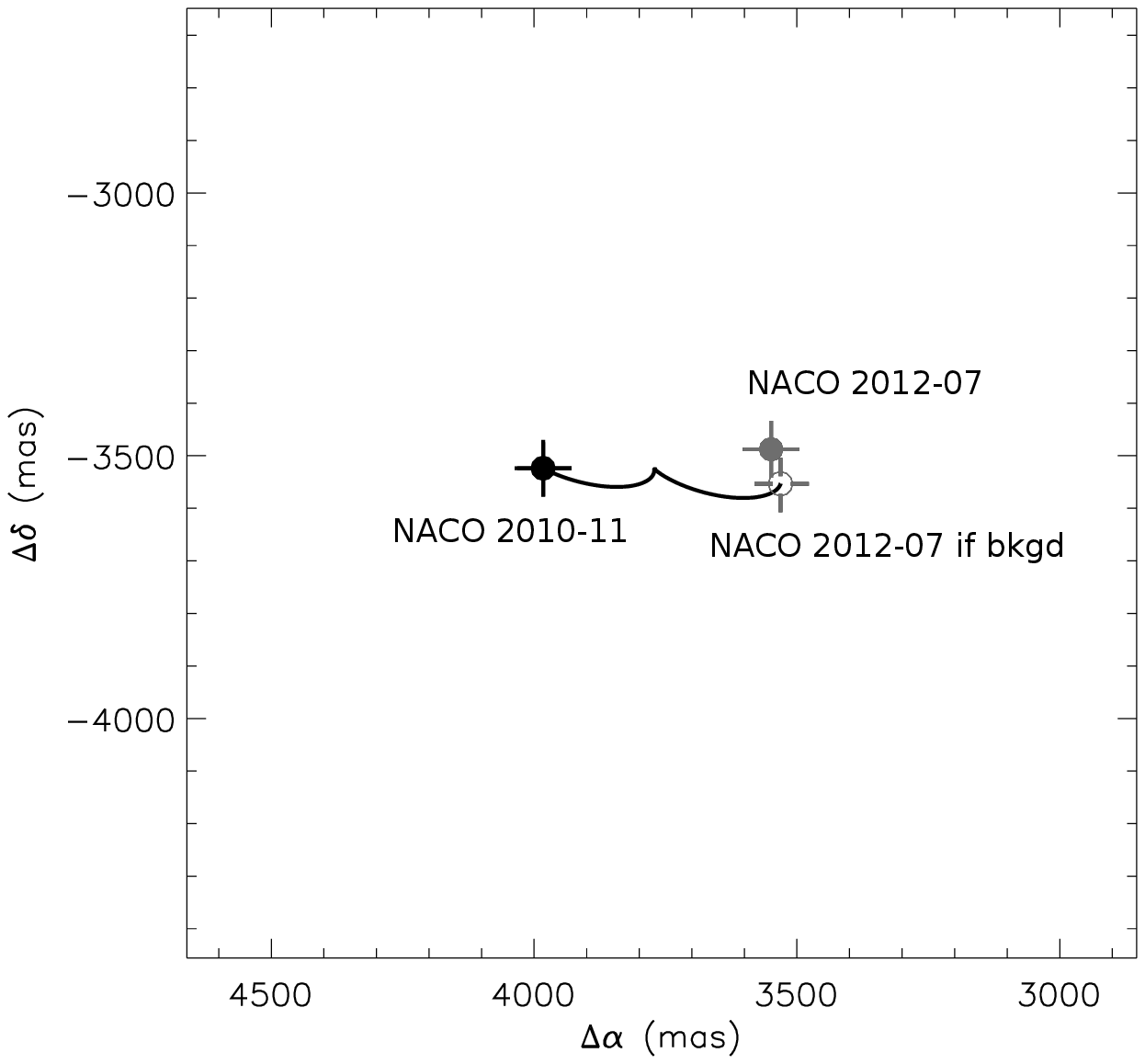}{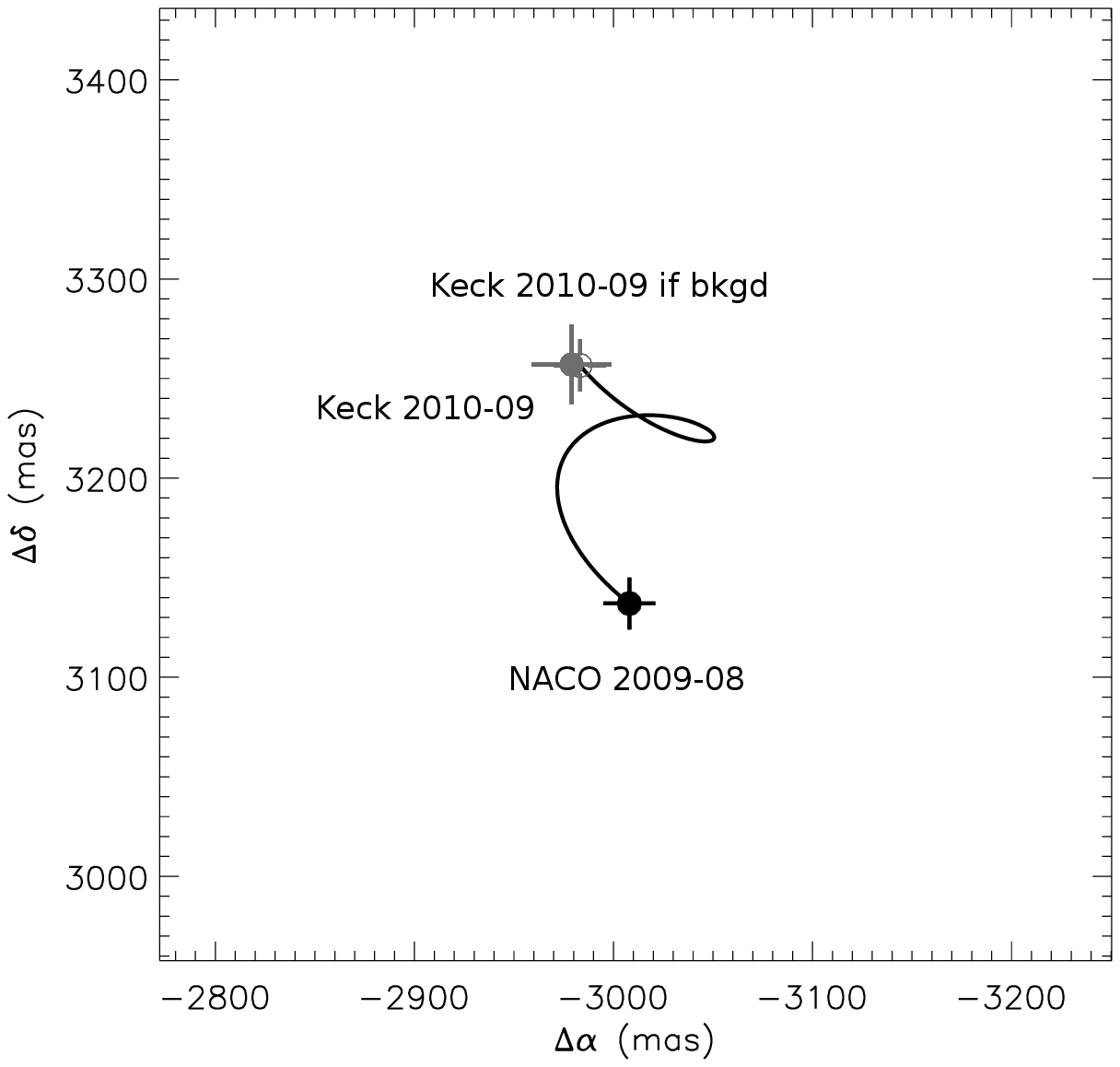}
 \caption{ Point sources with proper motion consistent with a background star (Top row Left: HIP 19893, Right: HIP 36515. Middle row Left: HIP 42333, Right: HIP 92919. Bottom row Left: HIP 105184, Right: HIP 108028).}
 \label{fig:background}
 \end{figure*}

We obtained follow-up observations for all point sources detected in our sample of Spitzer selected targets with VLT/NACO and Keck/NIRC2. A few of our sources were found to be likely binaries and thus were not considered for further analysis (see Table \ref{table:binaries} for the relative astrometry). The candidate companion around HD 73633 was resolved to be a binary. Six stars showed point sources (HIP 19893, HD 59967, HD 73350, HD 175742, HD 202628, HD 108028; see Figure \ref{fig:background}). We confirmed with second-epoch astrometry that all the point sources are not consistent with sharing common proper motion with their host stars, and thus are likely background stars (orbital motion for these widely separated point sources is negligible). The second-epoch data were obtained as part of this program with VLT/NACO, Keck/NIRC2 (HIP 36515, HIP 42333, HIP 92919, ), or archival data (VLT/NACO 088.C-0832 for HIP 19893, 089.C-0494 for HIP 105184, and 383.C-0600 for HIP 108028). Table \ref{table:astrometry_bg} lists the astrometry of these background objects.

\begin{deluxetable}{l c c c c c}
\tabletypesize{\scriptsize}
\tablecaption{Astrometry of background objects}
\tablehead{\colhead{Name} & \colhead{Epoch} & \colhead{RA (mas)} & \colhead{$\sigma_{RA}$} &\colhead{Dec (mas)} & \colhead{$\sigma_{Dec}$}}
\startdata
HIP 19893       & 2011.15 & -7142.1 & 27.0 & -3959.0 & 27.0 \\
                & 2012.13 & -7214.9 & 27.0 & -4165.5 & 27.0 \\
HIP 36515       & 2010.39 & 5541.9 & 54.0 & -4666.7 & 54.0 \\
                & 2010.98 & 5508.6 & 54.0 & -4589.6 & 54.0 \\
                & 2011.18 & 5636.1 & 30.0 & -4580.4 & 30.0 \\
HIP 42333       & 2010.31 & 2352.7 & 50.0 &  5334.2 & 50.0 \\
                & 2011.18 & 2660.0 & 20.0 & 5266.0 & 20.0 \\ 
HIP 92919       & 2010.82 & 1907.0 & 20.0 & 1809.7 & 20.0  \\
                & 2012.74 & 1637.1 & 20.0 & 2320.7 & 20.0 \\
HIP 105184      & 2010.97 & 3983.0 & 54.0 & -3523.8 & 54.0 \\
                & 2012.62 & 3549.1 & 54.0 & -3487.7 & 54.0 \\
HIP 108028      & 2009.74 & -3008.0 & 13.0  & 3137.0 & 13.0 \\
                & 2010.82 & -2979.0 & 20.0 & 3257.0 & 20.0 
\enddata
\label{table:astrometry_bg}
\end{deluxetable}

Several targets from the published high contrast imaging surveys we included in our complete debris disk and diskless sample have single epoch point source detections. The relative motions of these point sources have not been measured and thus it is not known if these are bound or  background objects. For these targets, we follow the conservative strategy in \citet{Bowler16} and limit the contrast floor to 1 $\sigma$ above the brightest point source reported in the images. Thus, we effectively remove these single epoch detections from our data, in order to prevent their influence on our statistical analysis.

\subsection{System Sensitivity Maps}
\label{sec:sensitivity_maps}
Using a semi-analytical method similar to \citet{Brandt14}, we calculate values for $P_i(m,a)$ as follows: 

\begin{equation}
P_i(m,a)=\int_{0}^{2}ds\,p(s)\,p(m,D=a\times s)
\label{eq:prob_dist}
\end{equation}

\noindent
where $P_i(m,a)$ is the probability of detecting a companion of mass $m$ at semi-major axis (SMA) $a$ for a given system $i$, and $s$ is the ratio of the projected separation $D$ over $a$. We integrate over $s$ from $s=0$ to $s=2$, allowing for eccentric orbits which can cause projection effects of up to doubling the SMA. Following the approach in \citet{Brandt14}, $p(s)$ is empirically derived from an eccentricity distribution $p(e)$, uniform up to $e_{\rm max}=0.8$. $p(s)$ is well fit by a piecewise linear function:

\begin{equation}\label{pup_cent}
p(s)\approx \bigg\{
\begin{array}{ll}
1.3s &0\leq s \leq 1 \\
-\frac{35}{32}(s-\frac{9}{5})  &1 < s < 1.8 
\end{array}
\label{eq:ecc_dist}
\end{equation}

The term $p(m,D=a\times s)$ is the probability of detecting a companion of mass $m$ at the projected separation $D$, and can be computed analytically from the contrast curve $C_i(D)$ as follows:

\begin{equation}
p_i(m,D)=0.5+0.5\, erf \left[ \frac{\tau}{\sqrt{2}}\left(\frac{L(m)}{C_i(D)}-1\right) \right]
\label{eq:mass_proj}
\end{equation}

where $\tau$ is the detection threshold (here $\tau=5$), $L(m)$ is the luminosity of a companion of mass $m$, following the COND evolutionary model \citep{Chabrier00,Baraffe03}. $C_i(D)$ is the $1\sigma$ contrast curve of object $i$, function of projected separation $D$. Because the effective inner working angle of first generation surveys is usually large ($>5\lambda/D$, where $\lambda$ is the observing wavelength and $D$ is the telescope diameter), we chose not to correct for small sample statistics \citep{Mawet14}. For the sake of continuity and comparison with previous studies, we chose to use the COND evolutionary model from \citet{Baraffe03}. The COND03 models were used in order to allow direct comparison with previous analyses \citep{Janson13,Rameau13,Bowler16,Galicher16}. However, as noted by \citet{Bowler16}, the COND model is part of hot-start model family, which begin with arbitrarily large radii and oversimplified, idealized initial conditions. It ignores the effects of accretion and mass assembly. The COND model represents the most luminous and thus optimistic outcome. We included the age uncertainties by drawing 10 samples from the age distributions and generating a detection probability map for each age sample. We take the average of these maps to be the final detection probability maps for each target (see Section \ref{sec:completed} for discussion about the age distributions).


\subsection{Companion occurrence rate}
\label{sec:occurrence}

While radial velocity surveys have constrained the mass and semi-major axis (SMA) distributions of gas giant planets at small and intermediate separations \citep[i.e.,][]{Cumming08,Bryan16}, direct imaging surveys present a unique opportunity to constrain the occurrence of gas giant planets at wide separations. We determine the occurrence rate of substellar companions around our sample of debris disk stars following methods outlined in \citet{Bowler15}. In short, these survey results can be characterized as Bernoulli trials.  While the number of detections is simply the number of substellar companions detected in these surveys, the number of trials, i.e. the number of times we asked whether we had a detection or non-detection, is given by the sum of sensitivities over a range of mass and semi-major axis and over the sample of systems.  Here, the number of trials is given by the following equation:

\begin{equation}
n = \frac{\sum^{N_{t}}_{i=1}\sum^{N_a}_{j=1}\sum^{N_m}_{k=1}P_i(m_k,a_j)}{N_aN_m}
\label{eq:missed}
\end{equation}

\noindent where $N_t$ is the number of systems, $N_a$ is the number of grid points in the specified SMA range, $N_m$ is the number of grid points in the mass range.  For this equation we adopt a double power law distribution of the form $\mathrm{d}N/(\mathrm{d}\log m \mathrm{d}\log a) \propto m^{\alpha}a^{\beta}$, and assume logarithmically flat distributions of $m$ and $a$ with $\alpha$ and $\beta$ equal to 0.  We note that given the low number of companion detections at wide separations by direct imaging surveys, it is still unclear what distribution their masses and separations follow. 

Since these survey results can be characterized as Bernoulli trials, we can model the probability distribution of occurrence rates $f$ as a binomial distribution given by the following equation:

\begin{equation}
    P(f | n,k) = \frac{\Gamma(n+1)}{\Gamma(k+1)\Gamma(n-k+1)}f^k(1-f)^{n-k}(n+1)
\end{equation}

In this equation, $n$ is the number of trials, and $k$ is the number of successes, i.e. the number of detected planets in a given range of mass and SMA. Here we generalize the binomial distribution by generalizing the binomial coefficient using Gamma functions in order to account for non-integer trials. 


 \begin{figure*}
\epsscale{1.1}
 \plottwo{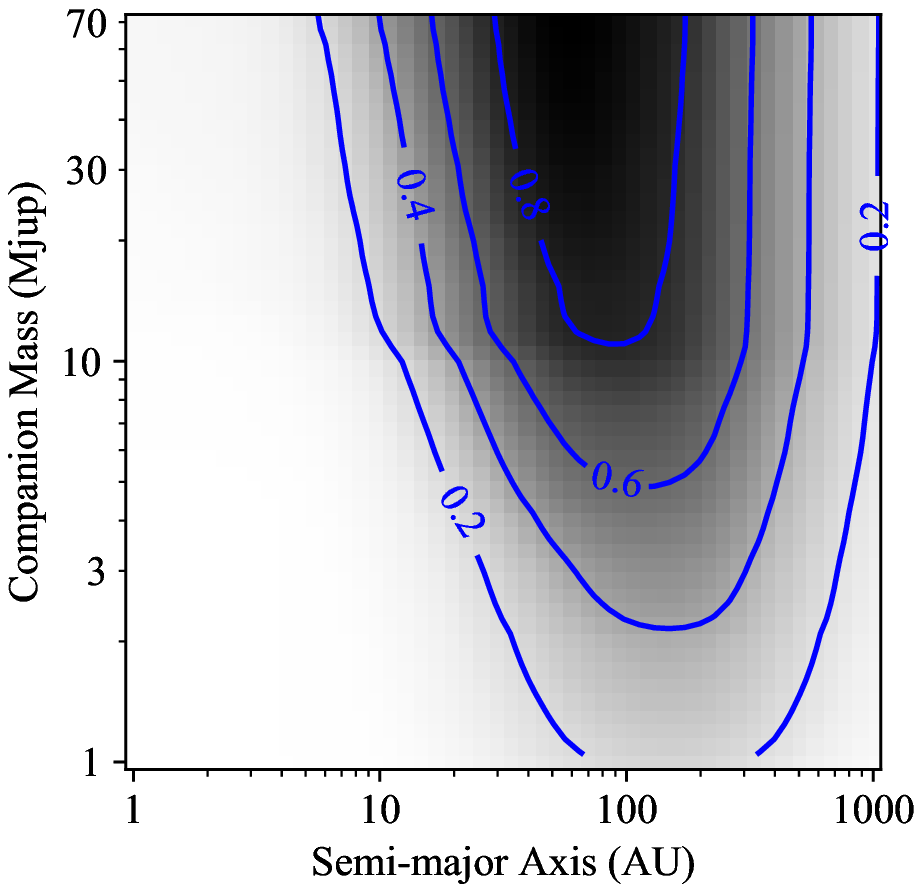}{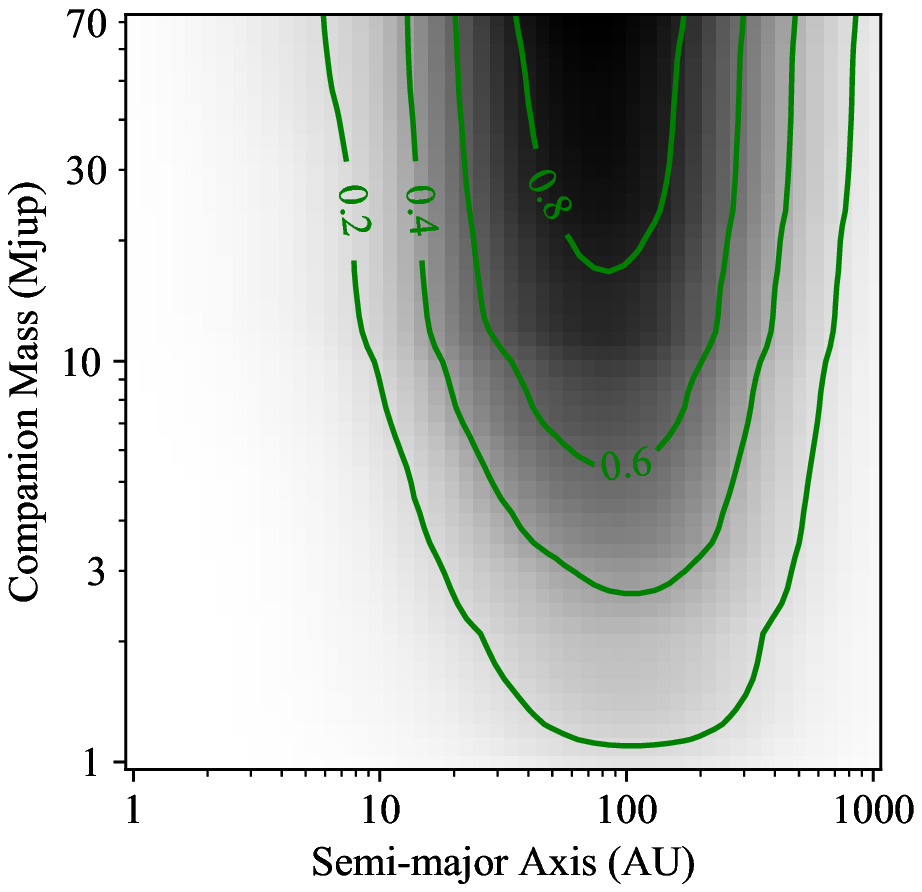}
 \caption{Detection probability map with contours showing the 20, 40, 60, and 80\% completeness for the debris disk sample (left) and the control sample (right) assuming hot-start and the \citet{Baraffe03} evolutionary model. Black indicates 100\% completeness.}
 \label{fig:common_average}
\end{figure*}

We calculate the occurrence rate for the debris disk sample over the mass range $5 - 20 M_{\rm Jup}$ and the SMA range 10 - 1000 AU, where we are relatively complete (see Figure \ref{fig:common_average}), which includes detections in systems HR 8799, HD 95086, $\beta$ Pic, and HD 106906. For the control sample of stars without debris disks, there was one reported companion, AB Pic B (see Section \ref{sec:results_pmcs}).

We find the occurrence rate of companions around stars with debris disks is 6.27\% with a 68\% confidence interval of 3.68 - 9.76\% for the range $5-20 M_{\rm Jup}$ and 10-1000 AU. For the control sample of diskless stars, the occurrence xfrate is 0.73\% with a 68\% confidence interval of 0.20 - 1.80\%. These distributions differ at the 88\% confidence level\footnote{In this work we list the confidence level rather than sigma since these posterior distributions are skewed and not Gaussian.}. We also calculated BIC values to compare these populations. We fit the combined distribution, the sum of the two binomial distributions, with one and two binomial distribution models, and calculated the BIC values from these model fits.  The two binomial distribution model was highly preferred ($\Delta BIC>10^{4}$) in comparison to the single binomial distribution model, suggesting that these two populations are drawn from different distributions. These results hint at a higher occurrence of giant planets around stars with debris disks than those without debris disks.
We note that although our statistical formalism handles the detection of only one companion around a star at a time, our inclusion of the HR 8799 four-planet system as one planetary system detection demonstrates that this occurrence rate applies to at least one companion per star.

\begin{deluxetable}{lcc}
\tabletypesize{\scriptsize}
\tablecaption{Occurrence Rates for Companions (5-20 $M_{\rm Jup}$ and 10-1000 AU) \\ at the 68\% confidence level (CL).}
\tablewidth{0pt}
\tablehead{\colhead{} & \colhead{Debris Disk} & \colhead{Control Sample}}
\startdata
Full Sample & 6.27\%, 68CL 3.68-9.76\%  & 0.73\%, 68CL 0.20-1.80\% \\
Early-type     & 9.94\%, 68CL 5.82-15.16\%  & --, 68CL 0-4.17\%      \\
Late-type & --, 68CL 0-4.61\%  & 2.18\%, 68CL 0.57-5.22\% 
\enddata
\label{table:occurrence}
\end{deluxetable}

\begin{figure}
\epsscale{1.1}
 \plotone{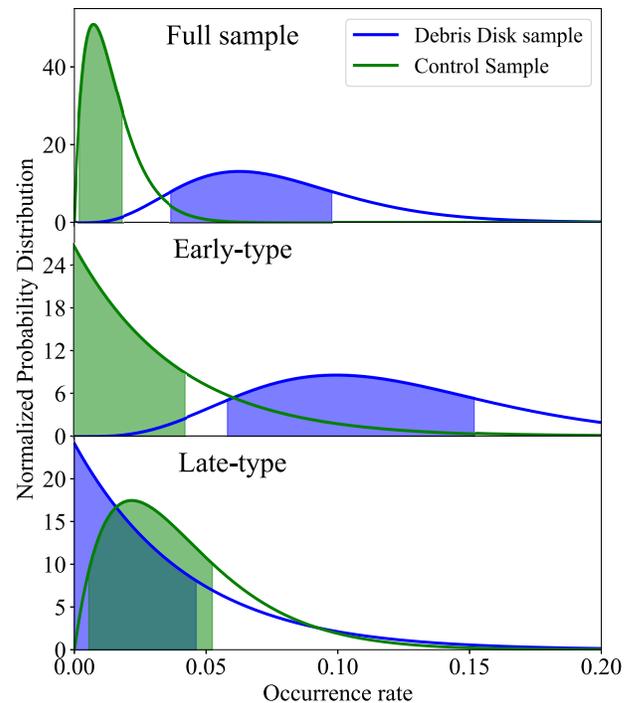}
 \caption{Normalized probability distributions for the debris disk and control samples for the full sample (top), early-type stars only (middle) and late-type stars only (bottom), with the 68\% confidence interval shaded.}
 \label{fig:distributions}
\end{figure}

We repeat these simulations with the early and late-type stars separately, to determine if the measured difference in occurrence rates among debris disk stars is the same or more prominent when considering only high-mass or low-mass stars. For the early-type stars, we find that an occurrence rate for the debris disk sample of 10.1\% with a 68\% confidence interval of 5.9 - 15.3\% and the control sample has a 68\% confidence level upper limit of 3.3\%, since there were no detections in this sub-sample. For the late-type stars only, we find an occurrence rate for the debris disk sample is a 68\% confidence level upper limit of 4.5\% and the control sample of 2.1\% with a 68\% confidence interval of 0.6 - 5.0\%. The early-type occurrence rates differ at the 83\% confidence interval, and the late-type occurrence rates are consistent at the 68\% level. Table \ref{table:occurrence} summarizes the occurrence rates for the debris disk and control samples, including the full sample of stellar types, as well as sub-samples of early- and late-type stars. The listed rate is the maximum of the probability distribution. \autoref{fig:distributions} shows the probability distributions comparing the debris disk sample and the control sample with the 68\% confidence interval shaded. We also calculated these occurrence rates using the power law distribution from \citet{Clanton16} using the Monte Carlo technique. We found these general occurrence rate trends to be consistent, and thus these results do not depend strongly on our choice of assumptions or priors.

Finally, we perform simulations on the control sample alone to ensure that it is not skewed by the large number of M stars (\autoref{fig:sp_type}). We repeat the simulations for the control sample without the M stars. The occurrence rate for the control sample without the M-stars is 1.2\% with a 68\% confidence interval of 0.3 - 2.8\%. The non-M star control sample and the control sample occurrence rates are consistent at the 68\% level, thus we can conclude that the disproportionate number of M stars in the control sample in comparison to the disk sample does not bias the derived occurrence rates. When the control sample without M-stars is compared with the debris disk sample, the samples still do not overlap at the 68\%. Our test shows that the control sample is minimally biased by the large number of low mass stars.

\section{Discussion}
\label{sec:discussion}
We performed our analysis on the complete sample of stars, as well as early- and late-type stars separately. For the complete sample of stars, our simulations hint at a higher occurrence rate for giant planets around stars with debris disks, as these samples differ at the 88\% confidence level. When considering only the early-type stars, we also find a higher occurrence rate of giant planets around stars with debris disks (77\%). This is consistent with predictions \citep{Zuckerman04,Wyatt05} that debris disks are the products of giant planets stirring and causing dust collisions, though these results suggest the need for more data to have stronger significance.

The structure and replenishment of a debris disk is often attributed to an eccentric perturbing planet, interior to the debris belt \citep{Wyatt05,Chiang09,Boley12,Nesvold15}. \citet{Nesvold16} considered the scenario of a debris disk being shaped by a perturbing planet external to the debris disk. An inclined planet can excite the disk eccentricities with the Kozai-Lidov mechanism. This suggests that the companion responsible for maintaining a debris disk structure may be further separated from its star, and thus easier for direct imaging discoveries. Brown dwarf companions have been directly imaged orbiting exterior to debris disks (i.e., HR3549;  \citealt{Mawet15}). Out of the planetary mass companions ($<20 M_{\rm Jup}$) discussed in this work, only HD 106906b orbits external to its debris disk. \citet{Nesvold17} use collisional and dynamical simulations to model the interactions between the planet HD 106906 b and the debris disk. They find that the planet can be responsible for the disk shape, and thus may have formed in situ (external to the debris disk). More generally, \citet{Lee16} demonstrate through disk modeling how the interaction between a single eccentric planet could produce a variety of observed disk morphologies. Our results are consistent with the theory that debris disks are the result of perturbing companions exciting collisions between dust, however we have too few companion detections to conclude whether the debris disks in our sample are shaped by an internal or external perturber.

In this paper we limited our data to published surveys, which does not include the large, ongoing, second generation data from the GPI Exoplanet Survey \citep{Macintosh14} and the SPHERE GTO \citep{Beuzit13}, as these are not yet completed. One test for those survey results would be to repeat the analysis here, and to adjust the SMA radii range based on the debris disk gap location. This will contribute to our understanding of whether the planet perturbers are present more often around stars with debris, and are located inside the debris disks or external \citep{Nesvold16}. Below, we derive our warm ($\sim150$K) debris radius, based on SED fitting, as often inside of our coronagraphic inner working angle (see Section \ref{subsec:debris_disk_radii} and Table \ref{table:disk_properties}). Hence, we cannot perform this adjusted radius test, as we are limited to cold outer dust and by the inner working angle of the first generation direct imaging instruments.

\subsection{Comparison to previous results}
Although the evidence is building, the degree to which circumstellar debris disks are the tracers of exoplanetary systems is still an unresolved issue, despite being an active area of theoretical work \citep[e.g.][]{MoroMartin07,Krivov10}. Indeed, in an attempt to understand the correlation between planets and debris disks, exoplanet host stars have been prime targets for space-based infrared observatories such as {\it Spitzer} and {\it Herschel} \citep[e.g.][]{Liseau10, DodsonRobinson11b}. However, studies such as these that use large samples from both populations have not produced statistically strong correlations between the two \citep{Bryden09}. 

The availability of the WISE all sky survey \citep{Wright10} has made it possible to search for correlations between the Kepler transiting systems and the presence of warm debris dust \citep{Krivov11,Ribas12,Lawler12}. For example, a transiting planet was discovered orbiting a 5-10 Myr star with a circumstellar disk \citep{David16, Mann16}. However, the frequency of debris disks in transiting systems is found to be only a few percent. We note that exoplanet host stars for transiting and radial velocity planets are on average older than the targets in our directly imaged sample. From an opposite approach, \citet{Morales12} use WISE to explore the incidence of warm (12 and/or 22 $\mu m$) dust around planet-host stars (independent of planet detection technique), and also found a $\sim 1\%$ excess incidence for main-sequence stars for the WISE detection limits. 

On the other hand, those debris disk systems that have significant spectral coverage into the thermal infrared, facilitating detailed SED fits, occasionally show evidence for dust belts organized into structures with both warm and cold thermal dust components \citep[e.g.][]{Kalas05, MoroMartin10, Morales11, Chen14, Ballering14}. While this is not definitive proof that planetary mass companions are present in the dusty systems, these dynamical structures provide a tantalizing hint of massive bodies responsible for debris sculpting. The work presented in this paper is the most comprehensive step in demonstrating this connection between wide and massive planetary companions and circumstellar debris. 

Comparing occurrence rates from other analyses is challenging due to different assumptions made in the occurrence rate calculations, SMA and mass ranges, as well as varying completeness achieved in the data. For our analysis, we chose a conservative mass range where we were most complete (see \autoref{fig:common_average}). Bearing these caveats in mind, we compare our occurrence rate for giant planets around stars with debris disks (6.2\% with a 68\% confidence interval of 3.6 - 9.7\%) to the large debris disk-selected surveys included in this paper. Our results are consistent with \citet{Wahhaj13}, who measure an upper limit occurrence rate of $20\%$ (68\% confidence level). \citet{Rameau13} find an occurrence rate of 10.8-24.8\% (68\% confidence). The higher occurrence rate found in \citet{Rameau13} may be the result of the smaller sample size and the large selected SMA and mass ranges (1-1000 AU and 1-13 $M_{\rm Jup}$) where the data are less complete.

\subsection{Debris Disk Radii Estimates}
\label{subsec:debris_disk_radii}
In order to place the sensitivity curves derived above in a more physical context, we have calculated approximate disk radii from the observed dust temperature, or temperature upper limit \citep[e.g.][]{Chen05,Hillenbrand08}. For the non-detections in our paper, these disk radii estimates combined with our contrast limits allow us to constrain the maximum mass of objects that could be present near the disk. 

We use the {\it Spitzer} photometry to estimate the disk temperatures and disk radii. Many of our targets show only infrared excesses at 24 or 70 $\mu$m. While this is sufficient to infer the presence of a debris disk, fitting the SED results in only a temperature upper limit and a radius lower limit. For those stars that show infrared excesses at both 24 and 70 $\mu$m, we fit a single temperature black body to these excesses, from which we derive a dust temperature. Table~\ref{table:disk_properties} shows several of the parameters derived for the sample including the dust temperature, disk radius, and the corresponding Jupiter mass limit for a companion at that radius, based on our contrast curves.

\begin{deluxetable}{l c c c c}
\tabletypesize{\scriptsize}
\tablecaption{Disk properties for targets with 24 and 70$\mu m$ excesses}
\tablehead{\colhead{Target} & \colhead{T$_{dust}$(K)} & \colhead{R$_{disk}$(au)} & \colhead{R$_{disk}$ ('')} &\colhead{$M_{\rm Jup}$ at R$_{disk}$}}
\startdata
HIP 7576         & 60     & 49 & 2.0 & 8.65 \\ 
HIP 36827        	& 114   &	5      & 0.2 & 34.5\\ 
HIP 42333        	& 45   & 150 & 6.3	 & 7.67 \\ 
HIP 74702        	& 49   & 80 & 5.0 & 3.1  
\enddata
\label{table:disk_properties}
\end{deluxetable}

With a derived dust temperature, the disk radius can quickly be calculated.  All the estimates for the disk sizes were constructed under the assumption of non-blackbody grains.  We used astronomical silicate properties to calculate the grain emissivity and equilibrium temperature assuming each star has the main sequence luminosity appropriate to the spectral types. We assumed the following sizes based on spectral type for the smallest grains, $a_{min}$, in the dust size distribution for our calculations: A0 is 5$\mu m$, F0 is 4$\mu m$, G0 is 3$\mu m$, K0 is 2$\mu m$, and M0 is 1$\mu m$.  These sizes are approximations anchored in previous modeling by members of our team of the Spectral Energy Distributions of debris disks spatially resolved by the Hubble Space Telescope \citep[e.g.][]{Krist10, Golimowski11}.  These assumptions allow us to convert observed temperatures into disk sizes. The disk inner radius always dominates the far-infrared emission, so the disk radii in Table~\ref{table:disk_properties} should be thought of as the disk inner edge.

\subsection{Theoretical explanation}

Over the last two decades, evidence has been marshaled in support of core-nucleated accretion \citep{Stevenson82,Pollack96} as a dominant formation mode of giant planets that reside in close proximity to their host stars \citep[e.g.][]{Fischer05, Miller11, Batygin16}. In contrast, the primary formation channel of more massive, distant bodies continues to be somewhat uncertain, since direct gravitational collapse \citep{Boss97} remains a distinct possibility at large stello-centric radii, where the natal gaseous nebulae would have been comparatively colder. In this regard, the preference for dusty debris disks to be accompanied by distant giant planets reported herein, points to the presence of refractory material as a marker for giant planet formation. 

The young ages of the host stars within our sample open a unique window into the primordial state of planetary systems that host long-period giant planets. In particular, the orbital architectures provide key extrasolar context for the early dynamical evolution of the solar system itself. The detailed orbital structure of the Kuiper belt \citep{Levison08} implies that the outer members of the solar system once occupied a much more compact (probably resonant) configuration, and were surrounded by a $\sim$20-50 $M_{E}$ debris disk that extended to $\sim$30 AU \citep{Tsiganis05,Nesvorny15}. Accordingly, the systems redetected by our survey likely represent the closest analogs to the young solar system within the currently known extrasolar planetary census. 

A closely related point follows regarding the typical evolutionary sequences of giant planet systems. It is generally established that planets should emerge from their protoplanetary disks on nearly circular, co-planar orbits. Subsequently, a large fraction of the giant planet sub-population evolves onto unstable trajectories, allowing planet-planet scattering to ensue, and shape the final orbital distribution \citep{Rasio96b, Beauge12}. Although this narrative reproduces the observed (RV) eccentricity distribution well \citep{Juric08}, the generic physical process that triggers the dynamical instabilities remains unclear \citep{Lega13}. To this end, within the framework of the Nice model, angular momentum exchange between the solar system’s outer planets and its primordial debris disk is invoked to initialize the transient instability \citep{Tsiganis05, Levison11}. Accordingly, the results reported herein provide the suggestion towards the potential universality of interactions between planets and debris disks as a mechanism responsible for igniting large-scale dynamical instabilities in planetary systems.

\section{Conclusions}
\label{sec:conclusions}

We describe a survey of stars with Spitzer-identified debris disks searching for directly imaged planets. We observed these targets with NIRC2/Keck and NACO/VLT and obtained follow-up data to confirm that all point sources in our data are consistent with background sources. We combined these results with the published contrast curves from four imaging surveys which directly target stars with debris disks: \citet{Wahhaj13,Janson13,Rameau13,Meshkat15}. Taking into account duplicates between the surveys, our sample of stars with debris disks includes 130 stars, 4 of which have planet detections (HR 8799, $\beta$ Pic, HD 95086, HD 106906). This is the largest unbiased sample of debris disks surveyed for long-period planets to date. In order to assess the occurrence rate of giant planets around stars with debris disks, we also obtained published contrast curves of 277 stars which do not have a debris disk, to act as a control sample. We verified that the age of the control sample is consistent with that of the debris disk sample, so as not to bias the results if planetary orbits evolve over time. We assume our sample of gas giant planets are distributed in mass and SMA space according to the double power law f(m,a)=$Cm^{\alpha}a^{\beta}$). Taking this companion distribution and our survey completeness in account, we find that the occurrence rate of giant planets around stars with debris disks is 6.27\% (68\% confidence interval 3.68 - 9.76\%), compared to 0.73\% (68\% confidence interval 0.20 - 1.80\%). These distributions differ at the 88\% confidence level. We ran simulations with the samples divided into early- and late-type stars to compare occurrence rates as a function of stellar mass. Our results show that early-type stars also show giant planet occurrence rates higher than early-type stars without debris disks, differing at the 77\% confidence level. The late-type star populations are consistent at below the 68\% confidence level. We also ran simulations for the control sample alone without the M-star population, in order to check if the sample is biased by the larger number of M-stars. The occurrence rate for the control sample without the M-stars is consistent with the control sample including the M-stars, and thus the control sample is not biased. 

Our comparison of the occurrence rates of gas giant planets between debris disk systems and our control sample suggests a tentative correlation. However, these results are sensitive to the small number of detected planets, thus we need more planetary mass detections and better completeness in mass and SMA to determine if this trend is significant. This work represents the results from first generation instruments. Second generation instruments will thus be needed to better understand the correlation between giant planets and debris disks.

\acknowledgments
We thank the anonymous referee for helpful comments and suggestions that improved this paper.
This work was performed with support from the Exoplanetary Science Initiative at the Jet Propulsion Laboratory, California Institute of Technology, under contract with NASA.
Part of this work was carried out at the Jet Propulsion Laboratory, California Institute of Technology, under contract with NASA. 
This work was performed in part under contract with the Jet Propulsion Laboratory (JPL) funded by NASA through the Sagan Fellowship Program executed by the NASA Exoplanet Science Institute. Support for this work was provided by NASA through Hubble Fellowship grant \#HST-HF2-51369.001-A awarded by the Space Telescope Science Institute, which is operated by the Association of Universities for Research in Astronomy, Inc. for NASA, under contract NAS5-26555. 
The data presented herein were obtained at the W.M. Keck Observatory, which is operated as a scientific partnership among the California Institute of Technology, the University of California and NASA. The Observatory was made possible by the generous financial support of the W.M. Keck Foundation.  The Authors wish to recognize and acknowledge the very significant cultural role and reverence that the summit of Mauna Kea has always had within the indigenous Hawaiian community.  We are most fortunate to have the opportunity to conduct observations from this mountain. We thank Rahul Patel, Geoff Bryden, Patrick Lowrance, and Grant Kennedy for detailed discussion about AB Pic.

\bibliographystyle{apj}  

\newpage
\clearpage

\appendix

\input{control_sample_properties.tex}
\clearpage

\clearpage
\newpage

\end{document}

%% file: targets_spitzer_sample.tex
\LongTables
\begin{deluxetable*}{l c c c c c c c c c c c c c c c c}
\tabletypesize{\scriptsize}
\tablecaption{Targets observed in our Spitzer sample.}
\tablewidth{0pt}
\tablehead{\colhead{Target} & \colhead{Instrument} & \colhead{Filter} & \colhead{Dates (UT)} & \colhead{$T_{int}$ (min)} & \colhead{$N_{images}$} & \colhead{Rot($^{\circ}$)} & \colhead{$0\farcs25^{*}$} & \colhead{$0\farcs5$} &  \colhead{$0\farcs75$} & \colhead{$1\farcs0$} & \colhead{$2\farcs0$} & \colhead{$3\farcs0$} & \colhead{$4\farcs0$}  & \colhead{$5\farcs0$}}
\startdata
HIP 1368     & NIRC2    &$Kp$      & 2010 Sep 27              & 65.7   & 197 & 35.9  & 9.1 & 10.4 & 11.9& 12.9& 14.7& 14.8& 14.7&--	\\
HIP 1499     & NIRC2    &$Kp$      & 2010 Sep 27              & 30.0   & 90  & 25.3  & 8.9 & 9.9 & 11.5 & 12.6 & 14.7& 14.6& 14.7& -- \\	
HIP 1598     & NIRC2    &$Kp$     & 2010 Sep 27              & 47.0    & 94   & 21.6  & 8.6& 10.0& 11.6& 12.6& 14.6& 14.7& 14.6& --\\
HIP 4148     & NIRC2    &$Kp$      & 2010 Sep 27              & 40.7   & 122  & 22.4  & -- & 10.1 & 11.6& 12.6& 14.5& 14.7& 14.6& --	\\
HIP 5944     & NIRC2    &$Kp$      & 2010 Sep 26              & 30.0   & 90 & 24.2  & 9.1& 10.9& 12.4& 13.6& 15.4& 15.6& --& --\\
HIP 7576     & NIRC2    &$Kp$     & 2010 Sep 26              & 28.0    & 70  & 20.4  & 8.7& 10.6& 11.8& 13.0& 14.1& 14.3& --& --	\\
HIP 8497     & NIRC2    &$Kp$      & 2010 Sep 26              & 35.0   & 70  & 22.0  & 8.9& 10.7& 12.3& 13.3& 15.5& 15.9& --& -- \\
HIP 17439    & NIRC2    &$Kp$      & 2010 Sep 26              & 50.0   & 152 & 17.7  & 9.4& 10.3& 11.9& 13.0& 15.1& 15.2& 15.2& --	\\
HIP 19893    & NACO	  & $L'$       & 2010 Nov 21              & 12.5   & 150    & 22.4  & --& 9.0& 10.0& 10.4& 12.3& 12.2& 12.4& 12.5\\
HIP 25775    & NACO	  & $L'$       & 2010 Nov 21              & 12.5   & 150    & 46.4  & --& 8.1& 8.9& 8.8& 9.4& 9.6& 9.4& 9.5 \\
HIP 30503    & NIRC2     &$Kp$     & 2010 Sep 26              & 45.5   & 90 & 19.7  & 7.8& 9.6& 10.9& 11.8& 14.4& 14.8& 14.8& -- \\
HIP 30729    & NACO    & $L'$       & 2010 Nov 22             & 10.5   & 127   & 7.5   & --& 7.1& 7.6& 7.6& 8.2& 8.4& 8.5& 8.6  \\
HIP 32919    & NIRC2	    &$Kp$     & 2010 Sep 27           & 28.5   & 57  & 19.7  & 8.4& 10.6& 12.1& 13.2& 14.3& 14.4& 14.3& -- \\
HIP 36515    & NACO    & $L'$       & 2010 Nov 21             & 25.0   & 250   & 58.9  & --& 9.1& 9.9& 10.3& 11.1& 11.0& 11.0& 11.1\\
HIP 36827    & NACO    & $L'$       & 2010 Apr 21             & 24.0   & 288   & 22.7  & --& 9.1& 9.3& 9.7& 9.9& 10.1& 10.1& 10.1 \\
HIP 42333    & NACO    & $L'$       & 2010 Apr 21             & 14.0   & 168   & 19.7  &--& 11.8& 12.0& 12.0& 12.1& 12.0& 12.1& 12.1	\\
HIP 43534    & NIRC2     &$Kp$     & 2012 Feb 02              & 28.0   & 56  & 12.5  & 7.3& 9.2& 10.7& 12.2& 13.4& 13.5& --& --	\\
HIP 44295    & NACO   & $L'$       & 2010 Apr 19              & 24.6   & 296      &  27.7  & --& 6.9& 7.6& 7.9& 8.4& 9.0& 9.0& 9.2	\\
HIP 50384    & NIRC2     &$Kp$     & 2012 Feb 02              & 3.3    & 109 &23.3  & 7.4& 9.3& 10.5& 11.8& 13.2& 13.2& --& --	\\
HIP 58451    & NACO  & $L'$        & 2010 Apr 21              & 33.3   & 400    & 69.5 & --& 8.2& 9.4& 9.9& 10.4& 10.4& 10.5& 10.5 \\	
HIP 73633    & NACO   & $L'$       & 2010 Apr 22              & 14.6   & 176   & 19.75 & --& 9.6& 9.7& 9.7& 9.9& 10.1& 10.0& 10.2 \\	
HIP 74702    & NACO   & $L'$        & 2010 Apr 20              & 10.0  & 120    & 9.3   & --& 8.2& 9.4& 9.9& 10.3& 10.4& 10.5& 10.5	\\
HIP 77952    & NACO   & $L'$       & 2010 Apr 21              & 18.0   & 216  &  33.1  & --& 9.2& 11.0& 11.6& 13.7& 13.8& 13.9& 13.9 \\
HIP 85561    & NACO   & $L'$       & 2010 Apr 21              & 7.3    & 88  & 5.3   & --& 7.5& 8.3& 8.6& 8.8& 8.7& 8.7& 8.7 \\
HIP 92919    & NIRC2  &$Kp$     & 2010 Sep 26                 & 32.5      & 130  &  17.3  & 8.7& 10.4& 12.0& 13.0& 15.0& 15.2& --& --\\
HIP 102626    & NIRC2    &$Kp$         & 2010 Sep 27         & 43.3    &  65  & 15.5  & 8.9& 10.2& 11.5& 12.6& 14.0& 14.0& 14.0& --	\\
HIP 105184   & NACO   & $L'$       & 2010 Nov 21              & 13.8   & 138   & 15.5  & --& 6.3& 8.6& 9.5& 10.6& 10.7& 10.8& 10.8	\\
HIP 108028   & NIRC2	  &$Kp$      & 2010 Sep 26              & 21   & 70  & 16.0  & 8.8& 9.8& 11.5& 12.7& 14.7& 14.9& --& --\\
HIP 112190   & NIRC2     &$Kp$     & 2010 Sep 26              & 32.0   & 80  & 22.1  & 9.2& 10.8& 12.1& 13.3& 15.1& 15.1& --& -- \\
HIP 117779   & NIRC2	    &$Kp$    & 2010 Sep 26              & 21.3 & 80   & 26.6  & 9.4& 10.8& 12.3& 13.4& 14.5& 14.7& --& --\\
\enddata
\tablenotetext{*}{Contrast limits in delta mag achieved from $0\farcs25$ to $5\farcs0$.}
\label{table:data_our_sample}
\end{deluxetable*}

%% file: debris_disk_sample_properties.tex
\LongTables
\begin{deluxetable*}{lllccccccccc}
\tablewidth{0pc}
\tabletypesize{\footnotesize}
\tablecaption{Properties of our debris disk sample}
\tablecolumns{11}
\tablehead{\colhead{Target} & \colhead{R.A.} & \colhead{Dec.} & \colhead{Sp. Type}  &\colhead{Dist. (pc)} &\colhead{Age (Myr)}& \colhead{Age Ref}& \colhead{Assoc.} & \colhead{V (mag)}&\colhead{H (mag)} &\colhead{Survey}}
\startdata
HIP544      & 00 06 36.785 & +29 01 17.40 & K0V             & 13.7$\pm$1.0   & $200^{+100}_{-50}$                            & LS06; 1      &   HL       & 6.13$\pm$0.05  & 6.13$\pm$0.05   & R13    \\
 HIP560      & 00 06 50.087 & $-$23 06 27.14 & F3V          & 39.1$\pm$1.0   & $12^{+2}_{-2}$                                & Z04; 1        &  BP       & 6.17$\pm$0.01  & 6.173$\pm$0.01  & W13    \\
 HIP682      & 00 08 25.746 & +06 37 00.49 & G2V                 & 39.1$\pm$1.0   & $90^{+130}_{-65}$                             & A08; 6   &   --           & 7.59$\pm$0.01  & 7.59$\pm$0.01   & J13    \\
 HIP1079     & 00 13 26.664 & +27 02 38.46 & G2V                & 34.0$\pm$19.0 & $12^{+2}_{-2}$                                & Z04; 1     &  BP          & 8.7$\pm$0.01   & 8.7$\pm$0.01    & W13    \\
 HIP1368     & 00 17 06.375 & +40 56 53.87 & M0.5V              & 15.0$\pm$1.0   & $295^{+205}_{-205}$                           & C14; 5     &   --         & 9.0$\pm$0.02   & 9.0$\pm$0.02    & TW -- Keck   \\
 HIP1481     & 00 18 26.122 & $-$63 28 38.98 & F8V                 & 41.0$\pm$1.0   & $30^{+15}_{-15}$                              & Z04; 1    &  TH           & 7.46$\pm$0.01  & 7.46$\pm$0.01   & W13    \\
 HIP1499     & 00 18 41.867 & $-$08 03 10.80 & G3V                 & 23.2$\pm$1.0   & $4700^{+2900}_{-3930}$                        & C11; 2,7   &  --          & 6.46$\pm$0.05  & 6.46$\pm$0.05   & TW -- Keck   \\
 HIP1598     & 00 20 00.409 & +38 13 38.64 & G1V               & 24.8$\pm$1.0   & $2490^{+2650}_{-1800}$                        & C11; 2,7     &   --       & 7.07$\pm$0.05  & 7.07$\pm$0.05   & TW -- Keck   \\
\enddata
\label{table:targets}
\tablecomments{Age references: A08: \citet{Apai08}, C09: \citet{Carpenter09}, C11: \citet{Casagrande11}, C14: \citet{Chen14}, D15: \citet{David15}, De15: \citet{Desidera15}, G12: \citet{Gontcharov12}, I10: \citet{Isaacson10}, L07: \citet{Lafreniere07}, LS06: \citet{LopezSantiago06}, M06: \citet{Moor06}, M08: \citet{Mamajek08}, Me09: \citet{Metchev09}, M09: \citet{Moor09}, M10: \citet{Maldonado10}, M11: \citet{Moor11}, M12: \citet{Mamajek12}, M13: \citet{Meshkat13}, M15: \citet{Meshkat15}, N04: \citet{Nordstrom04}, P09: \citet{Plavchan09}, P12: \citet{Pecaut12}, R07: \citet{Rhee07}, R05: \citet{Rieke05}, S05: \citet{Saffe05}, TW: This work, V12: \citet{Vican12}, Z01: \citet{Zuckerman01}, Z04: \citet{Zuckerman04}, Z11: \citet{Zuckerman11}, Z12: \citet{Zuckerman12}. Age determination methods: (1) cluster/group membership, (2) isochrone fitting, (3), x-ray age correlation, (4) Calcium H\&K emission age correlation, (5) rotation age correlation, (6) chromospheric activity, (7) Stromgren photometry, (8) Lithium abundances, (9) Rieke disk-age correlation. Associations: HL: Hercules-Lyra, BP: Beta Pic moving group, TH: Tucana/Horlogium, ABD: AB Dor moving group, UM: Ursa Majoris moving group, G2: Great Austral Young Association, Ar:Argus, Pl: Pleiades moving group, US: Upper-Scorpius, IC: IC2391, Ca: Castor, Co: Columba. } 
\end{deluxetable*}

%% file: control_sample_properties.tex
\LongTables
\begin{deluxetable*}{l c c c c c c c c}
\tabletypesize{\scriptsize}
\setlength{ \tabcolsep } {.1cm} 
\tablewidth{0pt}
\tablecolumns{9}
\tablecaption{Properties of the disk-free control sample}
\tablehead{\colhead{Target} & \colhead{R.A.} & \colhead{Dec.} & \colhead{Sp. Type} &\colhead{Dist (pc)} & \colhead{Age (Myr)} &\colhead{V (mag)}& \colhead{H (mag)} &\colhead{Survey}}
\startdata
 HIP\_5191                & 01:06:26.10 &$-$14:17:46.0 & K1   & 47.3$\pm$1.0    & $150^{+50}_{-30}$                                                     & 9.52$\pm$0.05  & 7.43$\pm$0.03 & Biller13     \\
 CD-58\_553               & 02:42:33.00 &$-$57:39:37.0 & K5   & 50.0$\pm$1.0    & $45^{+4}_{-4}$                                                        & 11.0$\pm$0.07  & 7.97$\pm$0.04 & Biller13     \\
 HD\_19668                & 03:09:42.29 &$-$09:34:46.5  & G0   & 37.4$\pm$1.0    & $150^{+50}_{-30}$                                                     & 8.49$\pm$0.02  & 6.79$\pm$0.04 & Biller13     \\
 HIP\_24947               & 05:20:38.00 &$-$39:45:18.0 & F6   & 48.3$\pm$1.0    & $45^{+4}_{-4}$                                                        & 7.38$\pm$0.01  & 6.22$\pm$0.03 & Biller13     \\
 HIP\_25283               & 05:24:30.10 &$-$38:58:10.0 & K6   & 18.0$\pm$1.0    & $150^{+50}_{-30}$                                                     & 9.05$\pm$0.04  & 6.11$\pm$0.03 & Biller13     \\
 HIP\_25486               & 05:27:04.80 &$-$11:54:04.0  & F7   & 27.0$\pm$1.0    & $23^{+3}_{-3}$                                                        & 6.3$\pm$0.01   & 5.09$\pm$0.03 & Biller13     \\
 UY\_Pic                  & 05:36:56.80 &$-$47:57:52.9 & K0   & 25.1$\pm$1.0    & $150^{+50}_{-30}$                                                     & 7.68$\pm$0.06  & 5.93$\pm$0.04 & Biller13     \\
 BD-13\_1328              & 06:02:21.90 &$-$13:55:33.0 & K4   & 39.0$\pm$1.0    & $150^{+50}_{-30}$                                                     & 10.55$\pm$0.05 & 7.89$\pm$0.04 & Biller13     \\
\enddata
\tablecomments{ References: \citet{ Lafreniere07, Vigan12, Biller13, Nielsen13,Brandt14, Bowler15,Galicher16}
}
\end{deluxetable*}